# Breaking New Ground in Software Defect Prediction: Introducing Practical and Actionable Metrics with Superior Predictive Power for Enhanced Decision-Making

**A PREPRINT**

**Carlos Andrés Ramírez Cataño[1], Makoto Itoh[2]**
[1]Doctoral Program in Risk and Resilience Engineering, Graduate School of Science and Technology, University of Tsukuba, 305-8573 Japan
[2]Institute of Systems and Information Engineering, University of Tsukuba, 305-8573 Japan
[2]Center for Artificial Intelligence Research, University of Tsukuba, 305-8573 Japan
[2]Tsukuba Institute for Advanced Research, University of Tsukuba, 305-8573 Japan

Corresponding author: Carlos Andrés Ramírez Cataño (e-mail: carlos@css.risk.tsukuba.ac.jp).

**ABSTRACT** Software defect prediction using code metrics has been extensively researched over the past five decades. However, prediction harnessing non-software metrics is under-researched. Considering that the root cause of software defects is often attributed to human error, human factors theory might offer key forecasting metrics for actionable insights. This paper explores automated software defect prediction at the method level based on the developers' coding habits. First, we propose a framework for deciding the metrics to conduct predictions. Next, we compare the performance of our metrics to that of the code and commit history metrics shown by research to achieve the highest performance to date. Finally, we analyze the prediction importance of each metric. As a result of our analyses of twenty-one critical infrastructure large-scale open-source software projects, we have presented: (1) a human error-based framework with metrics useful for defect prediction at method level; (2) models using our proposed metrics achieve better average prediction performance than the state-of-the-art code metrics and history measures; (3) the prediction importance of all metrics distributes differently with each of the novel metrics having better average importance than code and history metrics; (4) the novel metrics dramatically enhance the explainability, practicality, and actionability of software defect prediction models, significantly advancing the field. We present a systematic approach to forecasting defect-prone software methods via a human error framework. This work empowers practitioners to act on predictions, empirically demonstrating how developer coding habits contribute to defects in software systems.

## I. INTRODUCTION

Considering all relevant context, Software Defect Prediction (SDP) models have garnered significant attention over the past five decades due to their potential to improve the quality and security posture of software systems and optimize the allocation of testing resources, thereby reducing development costs [1–4]. These models are usually based on various types of metrics (e.g. code metrics, history measures) that aim to classify code fragments as either having defects or being defect-free.

The quest for metrics that not only forecast outcomes with precision but also translate into actionable insights has remained a formidable challenge. Current models lack of actionable messages that can guide developers in taking appropriate actions [5–7]. Existing methodologies often falter in bridging the critical gap between theoretical accuracy and practical applicability. The actionability of these metrics is frequently compromised by a lack of specificity, yielding ambiguous results that hinder the translation of predictions into concrete interventions. This challenge is compounded by issues of irrelevance to business objectives, and a failure to account for crucial contextual factors. Consequently, the insights generated often lack clear impact and encounter organizational resistance, collectively impeding their effective adoption in real-world software development lifecycles.

In addition, prediction models based on human factors theory are rarely investigated. Humans have inherent fallibility and limitations, making human error inevitable, frequent and predictable [8–10]. Given the collaborative nature of the software development process, there is a possibility for human errors to occur, and this is widely recognized as the primary cause of software defects.

The use of metrics based on human factors theory possesses the potential to offer a more efficient resolution to the software quality predicament compared to a sole emphasis on code and history metrics. The knowledge gained from the analysis of human factors-based metrics could assist developers to modify their behaviors, processes or procedures to proactively prevent and mitigate these errors.

This paper diverges from prior research in four notable aspects. First, we introduce a novel approach to constructing predictors by leveraging human factors theory. To do this, our approach builds upon the human error taxonomy proposed by Anu et al. [11] as the scheme for classifying and categorizing different types of human errors. We define specific performance shaping factors (PSFs), which are influences on human performance [12] during software development tasks and are directly linked to the errors within our framework. This taxonomy is rooted in the taxonomy of human error developed by James Reason, a noted authority on human error [8,10]. By utilizing this framework, we aim to provide a structured and comprehensive analysis of human errors within the context of our study.

Second, we measure and compare the performance of the method level defect prediction by using a) human error metrics, b) the traditional code and history metrics, and c) both of them combined. We would like to know whether our proposed models based on human error metrics could achieve similar or better performance at defect prediction to that of the state of the art models.

Third, we analyze metric importance on method level defect prediction to understand the predictive strength of each one of the metrics. We further determine the average metric rankings across all projects.

Fourth, we examine the experimental outcomes alongside real-world examples of open-source software defects and vulnerabilities identified using the novel metrics. We analyze the results while highlighting key findings and discussing their implications for the explainability, practicality, and actionability of software defect prediction.

To address our research objectives, we conducted empirical experiments guided by the following research questions:

*RQ1: What is the performance of method-level bug prediction models using state-of-the-art code metrics and history measures, human factors-based metrics or both?*

This question investigates the efficacy of our proposed models, based on human error metrics, for method-level bug prediction.

*RQ2: Which metrics have the most importance on method-level bug prediction?*

This question investigates the predictive power of individual metrics for method-level bug prediction, providing empirical evidence to guide practitioners in metric selection for model construction.

*RQ3: Compared to existing approaches, how do the newly proposed metrics enhance or hinder the explainability, practicality, and actionability of forecasting models as perceived by domain experts?*

This question investigates whether the new metrics advance the explainability, practicality, and actionability of forecasting models.

The following findings have been acquired through empirical evaluation across twenty-one large-scale open-source projects: (1) Defect prediction metrics based on human factors theory can be leveraged to effectively predict bug-prone methods; (2) models using our human error metrics have better performance than code and history metrics models, and the performance of the human error metrics model alone is, in average, better than the average performance of models that include both, human error metrics and code and history metrics; (3) different metrics have different prediction importance, but the human error metrics have a much better importance, in average, ranking on top of all metrics; (4) addressing long-standing issues, our novel metrics not only achieve superior accuracy over current state-of-the-art metrics but also provide transparent and actionable predictions.

We contend that our work has made significant advancements in the sate of the art, as evidenced by the following contributions:

- This study provides empirical evidence that substantiates the significance of leveraging human factors theory in SDP models, giving a way for practitioners to build derived prediction models based on human factors theory.
- This research provides empirical evidence of the superior performance of human error metrics in method-level SDP models, compared to the state of the art models based on code and history metrics.
- This study analyzes the importance of human factors-based metrics in prediction model outputs, providing empirical evidence of their consistently high average importance rankings and demonstrating their robustness across diverse software projects, despite noise and minor data variations.
- Highlighting the critical role of human factors in defect causation, the research findings are further corroborated by real-world discoveries of defects and vulnerabilities in open source software systems. As the novel metrics quantify factors shaping human programming behavior, they inherently provide explanations for defect origins and guide concrete future prevention strategies. Collectively, this work addresses the longstanding challenges of explainability, practicality, and actionability in software defect prediction.

In the next section we review the concepts. In Section III we discuss related work. In Section IV we present the design of the approach. Section V provides an overview of our approach to evaluation, including the methods employed, the analysis conducted, and the corresponding results. Section VI discusses the threats to validity. Finally, Section VII presents the discussions and conclusion of this study.

## II. CONCEPTS

**Software Defect.** An incorrect or missing step, process, or data definition in a computer program. [13]

**Human Error (HE).** A mistake. A human action that produces an incorrect result. It is a result of a planned sequence of mental or physical activities failing to achieve its intended outcome [8, 13]. In the context of this research, HE is an erroneous human behavior that leads to the introduction of a software defect in the source code.

**Software Defect Prediction (SDP).** Refers to the process of using various techniques and models to forecast and estimate the likelihood of defects or bugs in software systems [1–4].

**Explainability.** Refers to the degree to which a human can comprehend the rationale, underlying mechanisms, and cause-and-effect relationships governing a system's decisions or predictions [14].

**Human factors.** Applied scientific discipline focused on optimizing the interaction between humans and systems to enhance overall system performance, safety, and user well-being [15–16]. It involves designing tasks, tools, and environments that are compatible with human capabilities and limitations, thereby minimizing errors and maximizing efficiency. In this paper, the terms "human factors" and "cognitive psychology" are used interchangeably. This reflects that our novel metrics are derived from human error theory, encompassing insights from both the broader applied field of human factors and the foundational understanding of mental processes provided by cognitive psychology.

**Performance Shaping Factors (PSFs).** Human errors are driven by context [12]. In the realm of human reliability analysis, Performance Shaping Factors (PSFs) are understood as the contextual conditions or elements that significantly influence and shape human performance during task execution [17]. These factors are critical because human errors are not random occurrences but are instead driven by the specific context in which tasks are performed. More precisely, PSFs represent the intricate interaction among individual characteristics, technological interfaces, and organizational influences that collectively contribute to the likelihood of human errors [12]. Within the scope of this work, emphasis is placed on how specific combinations of software development behaviors and their associated PSFs create an "error-forcing context." This context increases the propensity for unsafe actions, thereby introducing defects into software systems. Performance shaping factors are used in systematic assessment methods to identify contributors to human errors and to provide a basis for quantifying those contributors systematically [12].

## III. RELATED WORK

To date, no approaches have been identified in the literature that automatically forecast the location of likely defects within a program using actionable and practical metrics. While numerous predictive models have been proposed, primarily relying on traditional software metrics, there has been limited exploration of alternative metrics, particularly those derived from the field of cognitive psychology.

### A. HUMAN FACTORS IN SOFTWARE DEFECT PREDICTION

Studies leveraging human factors theory for defect prediction are limited. One such notable recent effort by Huang et al. [18] explored the theory that software defects arise from human errors in error-prone scenarios. Their methodology involved five human analysts who manually performed task analysis on software requirements in the requirements phase. This qualitative approach yielded precise predictions for the location and form of 7 (31.8%) of the 22 defect types ultimately discovered in the codebase. Leveraging human error research, Anu et al. [11] investigated software quality through a systematic literature review spanning software engineering and psychology domains. Their work culminated in the Human Error Taxonomy, which expands upon Reason's established Slips, Lapses, and Mistakes taxonomy by introducing granular error classifications. This taxonomy offers a structured methodology for understanding and preventing human errors, and the subsequent faults, inherent in human-centric software engineering processes such as requirements engineering.

## B. ADDRESSING THE GRAND CHALLENGES IN SOFTWARE DEFECT PREDICTION

In a study at Google, Lewis et al. [5] suggested that bug prediction algorithms, primarily validated in academic lab settings, lack evidence of real-world utility in guiding developers. An experiment at Google, deploying such an algorithm, revealed no significant changes in developer behavior. This experience highlights critical characteristics necessary for bug prediction algorithms to gain developer acceptance and genuinely influence code evaluation practices. In another more recent study, Li et al. [7] explored prospective research directions and challenges in defect prediction. The work proposed potential ways to enhance the practicality, explainability, and actionability of defect model predictions. The core discussion of the work was around challenges across data quality (labeling and privacy-preserving sharing), multi-feature fusion in metrics, and model construction, evaluation, and interpretation (emphasizing effort, explainability, and actionability).

The preceding studies indicate that the limited explainability of current defect prediction models constrains their utility in guiding developers toward transparent decision-making. This lack of transparency erodes trust in their predictions and recommendations, ultimately impeding their widespread adoption in software development. Our work contributes to the field by proposing novel models that leverage human error-based metrics derived from cognitive psychology. We experimentally analyze their performance, demonstrating their potential to overcome existing challenges.

## IV. STUDY DESIGN

This section outlines our approach to building, training, testing and evaluating SDP models. We start by introducing our human error-based framework. We then present the selection of human error metrics, code metrics and historical measures utilized in this study. Subsequently, we explain how labeling of methods with defects is done. Afterwards, we explain the supervised machine learning (ML) algorithm to construct our prediction models. Finally, we describe the performance assessment of these models and the importance of the metrics in predicting defects at the method level.

### A. A HUMAN ERROR-BASED FRAMEWORK

#### 1) THE HUMAN ERROR TAXONOMY

In order to construct our human error metrics for SDP, it is essential to establish a comprehensive human error taxonomy that delineates errors within the specific context of software development. This taxonomy will serve as the foundation for categorizing and classifying different types of errors. Once the human error taxonomy is established, the subsequent step involves defining common PSFs that are known to be linked to these identified errors.

In this research, we have adopted the human error taxonomy developed by Anu et al [11]. This taxonomy is grounded in the widely recognized classification of human errors established by James Reason, a renowned authority in the field of human error. Anu et al. formulated this taxonomy by conducting a comprehensive literature review encompassing the domains of software engineering and psychology. The taxonomy incorporates error categories derived from an extensive analysis of 38 studies, which were sourced from various scholarly outlets such as journals, conferences, workshops, and book chapters. These studies encompassed a broad range of research topics spanning disciplines such as computer science, software engineering, medical quality, healthcare information technology, cognitive psychology, human-computer interaction, safety and reliability sciences, and cognitive informatics. By incorporating this robust and diverse set of studies, this taxonomy provides a comprehensive framework for classifying and understanding human errors in our research context.

Table I shows the errors identified in the literature by Anu et al. Table II shows how these errors are classified under Reason's human error taxonomy.

TABLE I
HUMAN ERRORS IDENTIFIED IN THE LITERATURE

| Error # | Error Name |
| --- | --- |
| 1 | Problem representation error |
| 2 | Requirements engineering; people do not understand the problem |
| 3 | Assumptions in gray area |
| 4 | Wrong assumptions about stakeholder opinions |
| 5 | Lack of cohesion |
| 6 | Loss of information from stakeholders |
| 7 | Low understanding of each other's roles |
| 8 | Not having a clear demarcation between client and users |
| 9 | Mistaken belief that it is impossible to specify non-functional requirements in a verifiable form |
| 10 | Accidentally overlooking requirements |
| 11 | Inadequate Requirements Process |
| 12 | Mistaken assumptions about the problem space |
| 13 | Environment errors |
| 14 | Information Management errors |
| 15 | Lack of awareness of sources of requirements |
| 16 | Application errors |
| 17 | Developer did not understand some aspect of the product or process |
| 18 | User needs not well-understood or interpreted by different stakeholders |
| 19 | Lack of understanding of the system |
| 20 | Lack of system knowledge |
| 21 | Not understanding some parts of the problem domain |
| 22 | Misunderstandings caused by working simultaneously with several different software systems and domains |
| 23 | Misunderstanding of some aspect of the overall functionality of the system |
| 24 | Problem-Solution errors |
| 25 | Misunderstanding of problem solution processes |
| 26 | Semantic errors |
| 27 | Syntax errors |
| 28 | Clerical errors |
| 29 | Carelessness while documenting requirements |

List of errors identified by Anu et al [11].

## 2) PERFORMANCE SHAPING FACTORS

Performance shaping factors are influential variables that significantly impact the occurrence and severity of errors. The primary objective of this study is to investigate the feasibility and efficacy of incorporating metrics derived from human factors theory into SDP models. To achieve this goal, it is crucial to begin by focusing on a limited set of PSFs that are intuitively and theoretically associated with the errors listed in our previously defined taxonomy.

TABLE II
HUMAN ERROR TAXONOMY

| Reason's Taxonomy | Human Error Class | Human Errors (Table I) |
|---|---|---|
| Slips | Clerical errors | 28, 29 |
|  | Term substitution | 5 |
| Lapses | Loss of information from stakeholders | 6 |
|  | Accidentally overlooking requirements | 10 |
|  | Multiple terms for the same concept errors | 5 |
| Mistakes | Application Errors: knowledge-based plan is incomplete | 1, 2, 16-23 |
|  | Solution choice errors | 24, 25 |
|  | Syntax errors | 26, 27 |
|  | Wrong Assumptions: knowledge-based plan is wrong | 3, 4, 12 |
|  | Environment Errors: knowledge-based plan is incomplete | 13 |
|  | Information management Errors: knowledge-based plan is incomplete | 14 |
|  | Inappropriate communication based on incomplete/faulty understanding of roles: knowledge-based plan is wrong | 7 |
|  | Not having a clear distinction between clients and users | 8 |
|  | Mistaken belief that it is impossible to specify non-functional requirements: knowledge-based plan is incomplete | 9 |
|  | Inadequate requirements process errors | 11 |
|  | Lack of awareness of requirements sources: knowledge-based plan is incomplete | 15 |

Error classification under Reason's [8] taxonomy.

This study prioritizes PSFs that demonstrate substantial impact on human performance, thereby potentially influencing the occurrence of the errors within our designated taxonomy. Additionally, this research places special attention to PSFs metrics that can derived from metadata contained in version management systems such as Git. This approach offers several advantages. It enhances feasibility and automation capabilities, enabling the efficient extraction of relevant metrics from available software development repositories. In addition, by leveraging the vast amount of open-source projects that utilize Git, this research can benefit from a wealth of available data. This abundance of data allows for extensive testing and validation of the proposed approach, providing a robust foundation for drawing meaningful conclusions. By focusing on performance factors obtainable from Git metadata, this study facilitates a practical and data-rich exploration of the relationship between human factors and software defects.

**Selection of Performance Shaping Factors.** In accordance with the human error taxonomy, knowledge-based mistakes are particularly dangerous because they occur in novel or unfamiliar situations where an individual must improvise and "think on their feet" with no established rules or procedures to guide them [8–10]. Consistent with this framework, the majority of errors identified in our study (24 out of 29) are classified within this category, as detailed in Table II. A careful examination of the specific human error classes in Table II reveals that most of these are attributable to an incomplete or faulty understanding of the problem space, as well as a lack of critical knowledge or information. Consequently, a primary focus for intervention should be on PSFs that either a) directly contribute to information or knowledge loss, or b) decrease an individual's capability to effectively execute a knowledge-based plan.

By taking into account these considerations and building upon existing research, we have made the decision to incorporate the following PSFs in this study:

**Memory Decay.** The gradual loss of information over time can significantly impact human performance in various activities. If a task has not been performed for a long time, there may be a need for relearning or refreshing the necessary skills and knowledge. This is known as the "learning decay" phenomenon, where skills and knowledge deteriorate over time without practice [19–23].

**Alertness.** An individual's state of being awake and responsive to stimuli can greatly influence human performance. Human performance tends to vary throughout the day due to factors like circadian rhythms, sleep-wake cycles, and natural fluctuations in alertness and energy levels [24–28]. For instance, individuals often experience a "morning peak" of alertness and focus, followed by an "afternoon dip" or "evening slump." While these patterns are common, it's important to note they may not reflect the behavior of all developers, as some are more productive at night (known as "night owls"). However, the probability function used in this research is based on a comprehensive study by Vallat et al. [28] grounded in data from a longitudinal study with 833 participants, and the calculation of "trait daytime alertness" for each individual provides a reliable basis for our metric, which uses the average alertness across all participants. It is important to note that the precise timestamps available in Git commits, which include the author's time zone, eliminate potential issues with time zone differences. In the future, this metric could be adapted to create personalized alertness profiles for different people.

### B. PREDICTION METRICS

**Code metrics and history measures.** Code metrics provide quantitative measures of the intrinsic properties of source code, while history metrics capture the evolution of code fragments as documented in the version control history. Both types of metrics have been widely used for predicting bugs [1–7], [29–32]. In this study, we employed thirteen code and history metrics recently analyzed in the exploratory study conducted by Mo et al. [29]. Our selection encompassed the top eight metrics identified as having the highest importance for method-level bug prediction, plus a few others that have been widely used in previous works. The chosen metrics and their reported rank are shown in Table III. This selection ensures that our evaluation framework remains robust and aligned with established standards in the field. By evaluating our newly proposed human error metrics against these established state-of-the-art metrics, we aim to conduct a comprehensive comparative analysis to assess the relative efficacy and predictive power of our metrics. These thirteen mentioned metrics can be calculated by parsing source code files, they have been extensively employed in bug prediction, with numerous studies substantiating their effectiveness [29–32].

**Human Error metrics.** We propose two human error-based metrics: *Memory Decay* and *Alertness*. These are based on the previously described PSFs with the same names.

Table IV enumerates the complete set of metrics employed in this research, with the first column providing a unique index for experimental identification. This collection comprises 13 state-of-the-art code and history metrics alongside the two novel HE-based metrics. This comprehensive collection constitutes the experimental metric set utilized in this study.

TABLE III
CHOSEN CODE AND HISTORY METRICS AND THEIR REPORTED PREDICTION IMPORTANCE RANK AMONG FORTY METRICS

| Metric Type | Metric | Rank |
|---|---|---|
| History | Number of authors | 1 |
| History | Added lines of code (LOC) | 2 |
| History | Changed LOC | 3 |
| History | Number of changes | 4 |
| History | Added LOC/LOC | 5 |
| History | Changed LOC/Number of changes | 7 |
| History | Added LOC/Deleted LOC | 8 |
| Code | Number of all lines | 10 |
| Code | Lines of code | 11 |
| History | Deleted LOC | 13 |
| History | Deleted LOC/LOC | 15 |
| Code | Number of blank lines | 28 |
| Code | Number of comment lines | 34 |

Code and history metrics analyzed by Mo et al [29].

### C. METRIC COMPUTATION

We developed an automated tool designed to calculate the required metrics by parsing the source code from the project's Git commit history within the temporal boundaries defined by a specified start date and a specified end date. The tool systematically analyzes the source code and the relevant metadata associated with each commit in the given time range, extracting pertinent data at the method level. Code and history metrics were calculated as described in previous research [29–32].

For each individual commit, HE-based metrics are systematically calculated and attributed to the corresponding commit author and logically linked to all methods updated in the commit. These data points are subsequently stored in a structured database for further analysis.

TABLE IV
LIST OF METHOD-LEVEL METRICS UTILIZED IN THIS STUDY

| Index | Metric Type | Metric |
|---|---|---|
| H1 | History | Number of authors |
| H2 | History | Added lines of code (LOC) |
| H3 | History | Changed LOC |
| H4 | History | Number of changes |
| H5 | History | Added LOC/LOC |
| H6 | History | Changed LOC/Number of changes |
| H7 | History | Added LOC/Deleted LOC |
| H8 | History | Deleted LOC |
| H9 | History | Deleted LOC/LOC |
| C1 | Code | Number of all lines |
| C2 | Code | Lines of code |
| C3 | Code | Number of blank lines |
| C4 | Code | Number of comment lines |
| E1 | Human Error | Memory decay |
| E2 | Human Error | Alertness |

List of code metrics, history measures and human error-based metrics.

**Memory Decay.** Each time an author commits changes to a specified method, a Memory Decay score is computed for each individual modification according to the Ebbinghaus forgetting curve [19,23] which describes the exponential rate at which information is forgotten over time, as presented in (1):

$$b = \frac{100k}{(\log(t))^c + k} \quad (1)$$

Here, b denotes "Savings", a measure of retention expressed as a percentage. This "Savings" quantifies the relative time preserved during a second learning trial due to prior exposure, with 100% indicating complete retention. The variable t represents time in minutes, commencing one minute before the conclusion of the initial learning phase. The constants c and k are empirically derived as 1.25 and 1.84, respectively.

The Memory Decay score for each method is calculated by summing the scores of all its associated changes. This total Memory Decay score is then fed into the machine learning model for defect prediction.

**Alertness.** Analogous to the computation of Memory Decay scores, each time an author commits changes to a method, an Alertness score is calculated for each individual modification. For each method, its total Alertness score is computed by summing the individual Alertness scores of all associated changes. The determination of this score relies on values derived as an approximation of the histogram of the number of alertness ratings as a function of time of day from Vallat et al. [28], as shown in table V:

TABLE V
DETERMINATION OF ALERTNESS SCORE

| Time Range | Alertness Score |
| --- | --- |
| Between 0:00 and 5:30 | 0.0 |
| Between 5:30 and 6:00 | 5.0 |
| Between 6:00 and 7:30 | 23.5 |
| Between 7:30 and 8:30 | 60.0 |
| Between 8:30 and 9:30 | 82.0 |
| Between 9:30 and 10:30 | 86.5 |
| Between 10:30 and 11:30 | 77.3 |
| Between 11:30 and 12:30 | 66.0 |
| Between 12:30 and 13:30 | 76.5 |
| Between 13:30 and 14:30 | 88.5 |
| Between 14:30 and 15:30 | 77.3 |
| Between 15:30 and 16:30 | 57.3 |
| Between 16:30 and 17:30 | 44.2 |
| Between 17:30 and 18:30 | 48.2 |
| Between 18:30 and 19:30 | 68.0 |
| Between 19:30 and 20:30 | 82.0 |
| Between 20:30 and 21:30 | 80.9 |
| Between 21:30 and 22:30 | 99.2 |
| Between 22:30 and 23:30 | 64.5 |
| Between 23:30 and 00:00 | 17.3 |

Determination of alertness score based on the time a code change is made.

### D. LABELING OF DEFECT PRONE METHODS

Consistent with prior research methodologies [29–32], we classified a method as defect-prone if it appeared in any commit identified as a bug fix at least once within the commit history. Conversely, methods that did not feature in any bug fix commits were categorized as not defect-prone.

### E. PREDICTION MODEL

#### 1) MACHINE LEARNING ALGORITHM

**Random Forest (RF).** Our forecasting model employs the Random Forest algorithm to generate method-level bug prediction models. Random Forest is an ensemble learning technique that constructs multiple decision trees during training and outputs the mode of their predictions for classification tasks. This algorithm is particularly advantageous for our forecasting task due to its ability to handle large datasets with high dimensionality and its robustness to overfitting. In this study, we conducted the defect prediction tasks utilizing the sklearn.ensemble module from Python's scikit-learn library, specifically configuring it to employ 100 decision trees (nestimators=100) within the Random Forest algorithm as done in previous studies [29–32]. The depth was set to None (max_depth=None), so the individual decision trees in the random forest are allowed to grow to their full potential.

#### 2) MODEL TRAINING AND VALIDATION

**Ten-fold cross-validation.** Our forecasting models employ the ten-fold cross-validation technique [33]. The ten-fold technique is a robust method for model evaluation that enhances the reliability of predictive performance metrics. In this approach the dataset is partitioned into ten equal subsets. Iteratively, each subset serves as a validation set while the remaining nine subsets collectively form the training set. This process is repeated ten times, with each subset being used exactly once as the validation set. The results are then averaged to produce a singular performance estimate, thereby mitigating the variability associated with data partitioning and providing a more generalized assessment of the model's efficacy.

#### 3) PERFORMANCE METRICS

The performance of our models was evaluated using three key metrics: the F1 score, the Matthew's Correlation Coefficient (MCC), and SHAP values.

**Accuracy, Recall, Precision, F1.** The F1 score is a widely employed metric in binary classification tasks, particularly within the domain of defect prediction, where it provides a balanced measure of a model's predictive performance. It is the harmonic mean of precision and recall, and it therefore penalizes models that perform poorly on either metric. Precision quantifies the proportion of positive predictions that were actually correct (i.e., the accuracy of predicting a defect), while recall measures the proportion of all actual positive instances that were correctly identified (i.e., the ability to find all defects). By integrating both measures into a single value, the F1 score offers a robust assessment of a model's ability to make accurate predictions while also identifying all relevant cases.

**Matthew's Correlation Coefficient (MCC).** This is a particularly informative and reliable metric for evaluating model performance on imbalanced datasets, a common scenario in defect prediction. Unlike other metrics that can be misleading on skewed data, the MCC considers all four values of the confusion matrix—true positives, true negatives, false positives, and false negatives. This comprehensive approach results in a more robust and honest measure of a classifier's performance. The MCC produces a value in the range of -1 to +1, where a value of +1 represents a perfect prediction, 0 signifies a prediction no better than random chance, and -1 indicates a complete inverse prediction. Consequently, a higher MCC value signifies a stronger and more reliable correlation between the predicted and actual classifications.

**Precision-Recall Area Under the Curve (PR AUC).** In this research we utilize PR AUC to evaluate the performance of our forecasting models. PR AUC is a common way to summarize a model's overall performance, and it is particularly better than Receiver Operating Characteristics (ROC) AUC for evaluating binary classifiers on imbalanced datasets [34]. Given that the majority of methods within the dataset are classified as not defect-prone, PR AUC emerges as the preferred metric for model evaluation.

**SHapley Additive exPlanations (SHAP Values).** Leveraging SHAP values, we calculate the importance rank of each prediction metric across the full dataset. Rooted in cooperative game theory, SHAP has transformed explainable AI by offering a unified approach to feature attribution [35–37]. These values assign the total model output to each feature, precisely quantifying its influence on a given prediction by considering all possible permutations, thereby ensuring a fair and consistent distribution of importance.

## V. EVALUATION
### A. RESEARCH QUESTIONS
This study investigates the following research questions:

> *RQ1: What is the performance of method-level bug prediction models using state-of-the-art code metrics and history measures, human factors-based metrics or both?*

This question investigated the efficacy of the code metrics and our proposed HE metrics for predicting defect-prone methods, and compared the performance of models utilizing each set of metrics individually and in combination.

> *RQ2: Which metrics have the most importance on method-level bug prediction?*

We aimed to identify the most influential metrics for bug prediction by extensively comparing the predictive power of all proposed metrics on a method-level. The findings offer empirical guidance for practitioners seeking to optimize metric selection for their prediction models.

> *RQ3: Compared to existing approaches, how do the newly proposed metrics enhance or hinder the explainability, practicality, and actionability of forecasting models as perceived by domain experts?*

We explore whether the new metrics improve the explainability, practicality, and actionability of forecasting models.

### B. SUBJECTS
Our empirical study was conducted on twenty-one diverse open-source projects, varying in size and domain. To ensure a representative selection of projects, we identified the top 21 C-language projects utilizing the Open Source Security Foundation's Criticality Score[1]. This initiative systematically tracks and assigns criticality scores to open-source software infrastructure projects, with higher scores indicating greater criticality. The target projects were: curl - network transfer library, emscripten - web compiler toolchain, esp-idf - IoT development framework, FreeRDP - Remote Desktop Protocol client, git - distributed version control, gnucash - personal finance software, libuv - async I/O library, linux - operating system kernel, linux-rbpi - Raspberry Pi Linux kernel, llvm-project - compiler infrastructure, obs-studio - video recording/streaming, openssl - SSL/TLS toolkit, openwrt - embedded Linux OS, optee_os - Trusted Execution Environment, php-src - PHP language source, qemu - processor emulator, redis - in-memory data store, RetroArch - emulator frontend, systemd - Linux init system, zfs - file system/volume manager, zephyr - RTOS for IoT.

Table VI summarizes the key characteristics of each project. It details the number of commits (ranging from 835 to 118,581), the number of files processed (345 to 38,626), the number of methods handled (2,073 to 203,808), the count of identified defective methods (533 to 22,529) and the total amount of lines of code contained in those methods (55,024 to 5,781,200) for which git data was studied. Given their significant contribution volumes, commit data for the Raspberry Pi kernel, the Linux kernel, and the LLVM Compiler suite were collected from January 1, 2024, through May 18, 2025. For all other projects, metadata extracion encompassed a wider range, from January 1, 2020, to May 18, 2025.

TABLE VI
RESEARCHED PROJECTS

| Project | Start Date | Commits | Files | Methods | Defective Methods | LOC |
|---|---|---|---|---|---|---|
| curl | 2020-01-01 | 5,189 | 1,123 | 7,576 | 2,210 | 249,080 |
| emscripten | 2020-01-01 | 5,986 | 10,451 | 30,544 | 2,585 | 449,091 |
| esp-idf | 2020-01-01 | 30,083 | 17,280 | 91,239 | 18,954 | 1,595,567 |
| FreeRDP | 2020-01-01 | 6,409 | 1,917 | 15,814 | 7,003 | 558,821 |
| git | 2020-01-01 | 8,493 | 981 | 10,560 | 1,870 | 277,531 |
| gnucash | 2020-01-01 | 3,114 | 1,163 | 8,418 | 758 | 211,200 |
| libuv | 2020-01-01 | 835 | 345 | 2,073 | 533 | 55,024 |
| linux-rbpi | 2024-01-01 | 118,581 | 33,366 | 203,808 | 22,529 | 5,781,200 |
| linux | 2024-01-01 | 80,250 | 33,122 | 201,053 | 22,299 | 5,709,394 |
| llvm-project | 2024-01-01 | 45,591 | 38,626 | 159,716 | 19,156 | 4,040,693 |
| obs-studio | 2020-01-01 | 5,108 | 1,904 | 26,387 | 1,727 | 527,456 |
| openssl | 2020-01-01 | 12,646 | 2,441 | 21,199 | 4,807 | 613,151 |
| openwrt | 2020-01-01 | 1,624 | 1,650 | 13,517 | 654 | 276,490 |
| optee_os | 2020-01-01 | 3,852 | 2,291 | 16,413 | 1,086 | 372,946 |
| php-src | 2020-01-01 | 11,530 | 2,492 | 27,877 | 4,162 | 4,318,750 |
| qemu | 2020-01-01 | 39,386 | 7,312 | 69,321 | 7,750 | 1,511,502 |
| redis | 2020-01-01 | 4,046 | 1,018 | 11,103 | 1,653 | 230,261 |
| RetroArch | 2020-01-01 | 9,972 | 4,346 | 29,437 | 5,255 | 1,020,833 |
| systemd | 2020-01-01 | 26,343 | 3,458 | 29,586 | 4,055 | 920,652 |
| zephyr | 2020-01-01 | 51,836 | 19,529 | 125,456 | 17,488 | 2,426,220 |
| zfs | 2020-01-01 | 4,840 | 1,250 | 10,284 | 1,497 | 331,826 |
| **Total** | | **475,714** | **186,065** | **1,111,381** | **148,031** | **31,477,688** |

The specific characteristics of the target projects.

An examination of the researched projects' attributes revealed several key insights. A critical observation is that only 13.3% of the methods were marked by developers as having received fixes, signaling the existence of a defect. This low percentage of labeled instances makes the dataset imbalanced, and poses a significant challenge for machine learning model training. Hence, as previously mentioned, the Precision-Recall Area Under the Curve (PR-AUC), F1 and MCC scores were utilized to evaluate model performance. The scope of this study involved the analysis of over one million methods, cumulatively exceeding 31 million lines of code.

---

1. Open Source Security Foundation's (OpenSSF) Criticality Score initiative: https://openssf.org/projects/criticality-score/

## C. RESULTS

***RQ1: What is the performance of method-level bug prediction models using state-of-the-art code metrics and history measures, human factors-based metrics or both?***

To provide an answer to this question, we employed the Random Forest algorithm to develop three types of bug prediction models: (a) Type1. Models based on code metrics and history measures, (b) Type2. Models based on HE metrics, and (c) Type3. Models incorporating both code and history metrics and HE-based metrics. Table VII presents the performance of each prediction model in terms of the PR AUC value. The fourth column displays the increment percentage of Type2 to Type1 models (2). Similarly the sixth column depicts the increment percentage of Type3 to Type2 models.

$$\frac{PR\ AUC_{Type2} - PR\ AUC_{Type1}}{PR\ AUC_{Type1}} \times 100\%$$

(2)

Project Curl provides a concrete example to demonstrate empirical findings. We observe that when utilizing traditional metrics or HE-based metrics separately, the corresponding PR AUC values for prediction models were 0.67 and 0.70, respectively. The Type2 model exhibits better performance than the Type1 model, with an improvement of 4.29%. The Type3 model, which incorporates both code metrics and HE-based metrics, further improves performance, attaining an PR AUC value of 0.73, indicating an increase of 5.12% compared to the Type2 model.

Upon calculating the average PR AUC values across all projects, we can observe intriguing results. The Type 2 model, which utilizes HE-based metrics, outperformed the Type 1 model in every single case. The Type 2 model has an average improvement of 38.41% compared to the Type 1 model. The Type 3 model, which combines both traditional metrics and HE-based metrics, experienced an average decrease in performance of -3.93%.

In addition to the PR AUC, it is important to measure a model's performance, focusing on its ability to make correct positive predictions. The predictive performance of the models is presented in Table VIII, which details the F1 scores and Matthew's Correlation Coefficients (MCC) for Type 1 model (based on code and history metrics) and Type 2 model (based on HE metrics). This comparison is a critical aspect of our findings, as it provides robust evidence of the predictive efficacy of each model approach. On average, the F1 score for models based on Type 2 metrics was 0.58, which represents a 45% increase over the average F1 score of 0.40 observed for models based on Type 1 metrics. Similarly, the average Matthew's Correlation Coefficient (MCC) for models based on Type 2 metrics was 0.52, representing a substantial 48.57% increase over the average MCC of 0.35 observed for models based on Type 1 metrics.

Notably, the consistent superiority of both the F1 and MCC values for the HE-based models reveals that this approach exhibits a demonstrably higher level of predictive capability and robustness compared to models built on traditional code and history metrics. This combined evidence allows for a confident conclusion that HE-based models are superior classifiers. Their higher F1 scores indicate a better balance between precision and recall, while the elevated MCC values confirm their enhanced reliability and effectiveness in handling the inherent challenges of imbalanced datasets often encountered in defect prediction.

By synthesizing the findings of all the analyzed projects as presented in Table VII and Table VIII, the following deductions can be extracted: First, through the implementation of HE-based metrics, it is possible to construct defect prediction models with satisfactory accuracy levels. Second, the predictive model grounded in cognitive psychology theory uniformly outperformed the alternatives, achieving higher scores for PR AUC, F1, and MCC across all evaluation scenarios. In terms of PR AUC performance, the models that draw on human factors data outdo the models that are based on code and history metrics, with gains of between 2.87% and 124.78%. Third, the combination of both traditional metrics and HE-based metrics, experienced an average decrease in performance with 15 out of the 21 projects experiencing a decrease between -1.28% and -20.00%.

**Answer to RQ1:** The average PR AUC values obtained for (a) code and history metrics: 0.49, (b) HE-based metrics: 0.65 and (c) models incorporating both code and history metrics with HE-based metrics: 0.62. Models based on human error metrics demonstrated a substantial improvement in predictive power, with an average of 38.41% enhancement over current state-of-the-art code and history metrics.

***RQ2: Which metrics have the most importance on method-level bug prediction?***

To provide an answer to this question, we evaluated metric prediction importance in model output by analyzing SHAP values with ten-fold cross-validation. The sum of the absolute values of the SHAP values quantify how much each feature contributes to the final prediction, considering all possible combinations of features.

The initial step involved the computation of the SHAP values for the proposed features. This process is simultaneously incorporated into the estimation of the PR AUC values. The features were subsequently arranged in ascending order, with a total of n features, where n denotes the number of features considered. The feature with the highest SHAP value was assigned a rank of 1 (best), while the feature with the lowest SHAP value was allocated a rank of 15 (worst). It is noteworthy that a metric with a higher rank signifies that it provides a greater impact on model output, thereby indicating that it is a more crucial attribute in the defect prediction process. The distribution of ranks for each metric is illustrated in Fig. 1, with each box plot showing two numbers that depict the range of ranks for the metric under consideration.

TABLE VII
PR-AUC VALUES OF THE DEVELOPED PREDICTION MODELS

| Project | Type 1 | Type 2 | ↑ (%) | Type 3 | ↑ (%) |
|---|---|---|---|---|---|
| curl | 0.67 | 0.70 | 4.29% | 0.73 | 5.12% |
| emscripten | 0.37 | 0.55 | 48.22% | 0.53 | -4.38% |
| esp-idf | 0.66 | 0.87 | 31.13% | 0.82 | -5.19% |
| FreeRDP | 0.82 | 0.87 | 5.99% | 0.88 | 0.84% |
| git | 0.46 | 0.48 | 3.80% | 0.49 | 3.35% |
| gnucash | 0.35 | 0.51 | 45.93% | 0.48 | -5.35% |
| libuv | 0.49 | 0.81 | 65.97% | 0.72 | -10.49% |
| linux-rbpi | 0.26 | 0.48 | 84.14% | 0.39 | -17.06% |
| linux | 0.35 | 0.54 | 57.08% | 0.48 | -11.40% |
| llvm-project | 0.30 | 0.68 | 124.78% | 0.55 | -20.00% |
| obs-studio | 0.49 | 0.58 | 17.89% | 0.58 | 0.23% |
| openssl | 0.63 | 0.72 | 14.62% | 0.71 | -1.28% |
| openwrt | 0.43 | 0.80 | 84.11% | 0.73 | -8.38% |
| optee_os | 0.41 | 0.60 | 45.82% | 0.56 | -7.84% |
| php-src | 0.62 | 0.72 | 14.95% | 0.70 | -1.82% |
| qemu | 0.39 | 0.51 | 28.48% | 0.51 | 0.93% |
| redis | 0.60 | 0.74 | 24.47% | 0.70 | -4.88% |
| RetroArch | 0.55 | 0.79 | 44.07% | 0.77 | -2.38% |
| systemd | 0.41 | 0.42 | 2.87% | 0.47 | 12.85% |
| zephyr | 0.52 | 0.65 | 25.71% | 0.64 | -2.12% |
| zfs | 0.50 | 0.66 | 32.20% | 0.64 | -3.21% |
| **Overall Average** | **0.49** | **0.65** | **38.41%** | **0.62** | **-3.93%** |

Precision-Recall Area Under the Curve Values.

TABLE VIII
RANDOM FOREST PERFORMANCE COMPARISON

| Project | Type 1 (F1) | Type 2 (F1) | Type 1 (MCC) | Type 2 (MCC) |
|---|---|---|---|---|
| curl | 0.58 | 0.64 | 0.44 | 0.50 |
| emscripten | 0.29 | 0.49 | 0.29 | 0.45 |
| esp-idf | 0.59 | 0.78 | 0.51 | 0.72 |
| FreeRDP | 0.75 | 0.83 | 0.55 | 0.69 |
| git | 0.36 | 0.44 | 0.30 | 0.34 |
| gnucash | 0.25 | 0.44 | 0.25 | 0.41 |
| libuv | 0.41 | 0.70 | 0.27 | 0.60 |
| linux-rbpi | 0.26 | 0.48 | 0.24 | 0.43 |
| linux | 0.16 | 0.41 | 0.15 | 0.37 |
| llvm-project | 0.19 | 0.59 | 0.19 | 0.55 |
| obs-studio | 0.45 | 0.53 | 0.44 | 0.51 |
| openssl | 0.55 | 0.66 | 0.45 | 0.56 |
| openwrt | 0.38 | 0.71 | 0.39 | 0.70 |
| optee_os | 0.31 | 0.51 | 0.31 | 0.49 |
| php-src | 0.53 | 0.64 | 0.47 | 0.58 |
| qemu | 0.29 | 0.46 | 0.28 | 0.41 |
| redis | 0.52 | 0.65 | 0.47 | 0.59 |
| RetroArch | 0.47 | 0.68 | 0.41 | 0.61 |
| systemd | 0.30 | 0.38 | 0.28 | 0.30 |
| zephyr | 0.44 | 0.57 | 0.39 | 0.51 |
| zfs | 0.41 | 0.56 | 0.36 | 0.50 |
| **Overall Average** | **0.40** | **0.58** | **0.35** | **0.52** |

Performance comparison for F1 and MCC values between models based on code metrics and models based on HE metrics.

As an example, examination of the first box diagram depicted in Fig. 1, it can be observed that the H1 metric (representing Number of authors) is assigned a rank ranging from 1 to 14 across all prediction models. In contrast, the E1 metric (representing Memory decay) is assigned a rank varying from 1 to 4 across all projects. In addition, calculating the average range (R = Maximum value – Minimum value) for the evaluated metrics we obtain: code metrics, 11.5; history measures, 10.44; and HE-based metrics, 4.00. Based on this analysis, the following inferences can be drawn: (1) The ranks assigned to each metric are variable across the studied projects; (2) the distribution of ranks for different metrics varies significantly across all projects. Human Error-based metric ranks exhibited substantially less variability than both code metrics and history measures; (3) Generally, the metrics based on cognitive psychology theory tend to rank higher than the traditional code and history metrics. Figure 2 shows prediction importance graph for the Curl project.

To provide additional insight into the significance of each metric, we calculated the average rank of each metric, as presented in Table IX. The E1 metric, which denotes Memory decay, obtained the highest average rank of 2.19, indicating that it carries the most weight in predicting defect-prone methods for the examined projects. Based on the presented information we can also find that the human error-based metrics have larger prediction power than the code and history metrics.

TABLE IX
METRIC RANKINGS

| Index | Metric Type | Metric | Rank |
|---|---|---|---|
| H1 | History | Number of authors | 4.90 |
| H2 | History | Added lines of code (LOC) | 11.24 |
| H3 | History | Changed LOC | 9.95 |
| H4 | History | Number of changes | 3.67 |
| H5 | History | Added LOC/LOC | 7.71 |
| H6 | History | Changed LOC/Number of changes | 9.57 |
| H7 | History | Added LOC/Deleted LOC | 9.76 |
| H8 | History | Deleted LOC | 5.62 |
| H9 | History | Deleted LOC/LOC | 6.48 |
| C1 | Code | Number of all lines | 8.71 |
| C2 | Code | Lines of code | 10.86 |
| C3 | Code | Number of blank lines | 11.76 |
| C4 | Code | Number of comment lines | 14.14 |
| E1 | Human Error | Memory decay | 2.19 |
| E2 | Human Error | Alertness | 3.14 |

Rank of metrics. A lower value (higher rank) is better.

**Answer to RQ2:** The predictive capabilities of metrics can vary at the method level. In accordance with our findings, the E1 metric, which refers to the Memory decay, has the highest significance for method-level defect prediction. Additionally, we observed that, in general, the HE-based metrics tend to exhibit greater predictive power than the code and history metrics. The importance rank distribution of HE-based metrics showed considerably less variability than both code metrics and history measures.

## VI. THREATS TO VALIDITY

**Internal validity.** The first threat to our study is the limited amount of Performance Shaping Factors-derived human error metrics that we explored. As a result of this limitation, we cannot generalize our findings to encompass all human error-based techniques in the field of software defect prediction. However, a quantitative analysis intended for screening human failure events generally does not necessitate an exhaustive inventory of PSFs [38]. Furthermore, a comprehensive list of PSFs may not enhance the precision of quantification beyond that achieved through a screening analysis. Ultimately, the critical consideration lies not in the quantity of PSFs, but rather in the purpose for which those PSFs are employed.

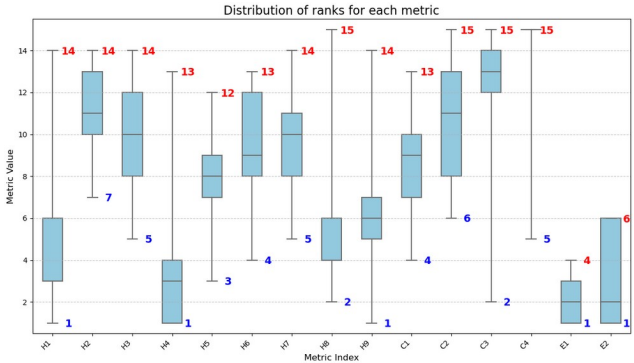

**Fig. 1.** Rank distribution for each metric.

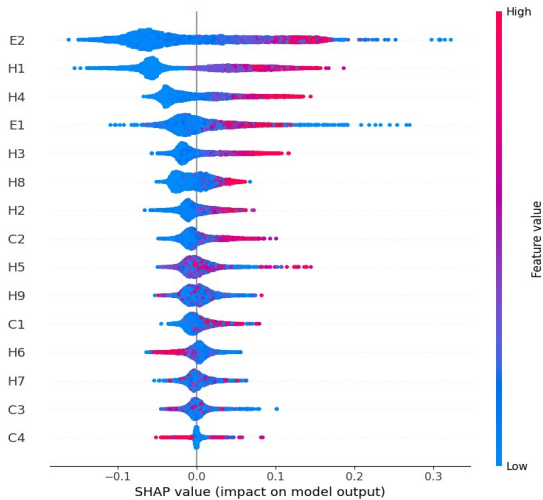

**Fig. 2.** Prediction importance of all features in curl.

However, in future studies, we plan to expand our investigation to include a wider range of techniques and compare their results against those obtained here. Additionally, in this paper, we have provided a comprehensive description of our methodology, experimental setup, and the dataset used. This enables other researchers to contribute to our study or to explore other unexplored techniques.

A second threat arises from the precision of bug data. Our approach involved leveraging the correlation between bugs and commit messages to identify commits that address bugs and subsequently label the methods prone to bugs. Nonetheless, it is essential to acknowledge that we cannot ensure complete impartiality in our findings for two reasons. Firstly, a method that is modified in a commit aimed at bug fixing may not necessarily be altered specifically to resolve a bug. Secondly, the absence of a direct and explicit connection between bug-fixing commits and their location in the revision history poses challenges in accurately pinpointing them.

**External validity.** While the datasets used in our work have been meticulously labeled based on ground truths, it is important to note that the number of datasets is limited. Consequently, it becomes challenging to generalize our results to other datasets and domains. Future work should focus on further investigating our study using additional datasets. Moreover, it is worth mentioning that all the projects used in our experiments are C programming language projects. Although these projects are widely utilized and popular projects, it is possible that our findings may not be fully applicable to projects implemented in different programming languages or domains.

## VII. DISCUSSION AND CONCLUSION

In this section we test our novel metrics on a different ML algorithm, discuss the experimental results and further support the validity of our HE-based metrics by utilizing our framework to find issues in real-world systems.

### A. THE COMPATIBILITY OF OUR APPROACH WITH OTHER MODELING PARADIGMS

In this study, we have demonstrated that our models, trained using the Random Forest algorithm, are capable of accurately identifying defect-prone methods. To mitigate the potential for inductive bias inherent in any single algorithm and to assess the broader applicability of our approach, we implemented a second, distinct machine learning model. For this purpose, we selected the LightGBM algorithm, a highly efficient gradient boosting framework developed by Microsoft research that has been widely and successfully applied in prediction tasks [39]. The choice of LightGBM was particularly motivated by the challenges posed by the massive repository sizes of projects like Linux and LLVM. Its efficiency stems from a histogram-based algorithm that bins continuous feature values, which not only significantly accelerates the training process but also requires substantially less memory. Furthermore, LightGBM's design for large-scale and high-dimensional data, along with its support for parallel and distributed computing, makes it exceptionally scalable for the extensive datasets examined in this research. The successful integration of our approach with LightGBM, a powerful alternative to Random Forest, helps to substantiate the versatility and robustness of our framework, suggesting its potential for application with a wide variety of modeling methods.

The LightGBM algorithm was configured for a binary classification task with the following hyperparameters. The model was optimized to address the class imbalance in the dataset and utilized the Area Under the Precision-Recall Curve (AUPRC) as its primary evaluation metric (metric='average_precision'). The specific parameters are detailed below:

    objective='binary',
    metric='average_precision',
    n_estimators=2000,
    learning_rate=0.05,
    num_leaves=31,
    is_unbalance=True.

The performance results for the LightGBM implementation are detailed in Table X and Table XI, which summarizes the F1 scores, Matthew's Correlation Coefficients (MCC), and Precision-Recall AUC for both Type 1 and Type 2 models. On average, models utilizing Type 2 metrics achieved an F1 score of 0.53, which represents a 15.21% increase over the average F1 score of 0.46 demonstrated by models based on Type 1 metrics. Similarly, a substantial performance advantage was observed with the MCC, where the average score for Type 2 metrics was 0.46, a 24.32% increase over the 0.37 average score of Type 1 metrics. These consistent results for both the F1 and MCC scores provide compelling evidence of the superior predictive capability and robustness of the models leveraging Type 2 metrics.

TABLE X
LIGHT-GBM PERFORMANCE COMPARISON

| Project | Type 1 (**F1**) | Type 2 (**F1**) | Type 1 (**MCC**) | Type 2 (**MCC**) |
|---|---|---|---|---|
| curl | 0.63 | 0.64 | 0.46 | 0.47 |
| emscripten | 0.37 | 0.40 | 0.34 | 0.39 |
| esp-idf | 0.59 | 0.69 | 0.48 | 0.62 |
| FreeRDP | 0.75 | 0.80 | 0.54 | 0.63 |
| git | 0.49 | 0.45 | 0.37 | 0.32 |
| gnucash | 0.38 | 0.38 | 0.32 | 0.33 |
| libuv | 0.47 | 0.69 | 0.23 | 0.57 |
| linux-rbpi | 0.38 | 0.43 | 0.33 | 0.37 |
| linux | 0.32 | 0.37 | 0.25 | 0.29 |
| llvm-project | 0.34 | 0.43 | 0.24 | 0.36 |
| obs-studio | 0.46 | 0.47 | 0.48 | 0.49 |
| openssl | 0.59 | 0.63 | 0.45 | 0.51 |
| openwrt | 0.16 | 0.52 | 0.19 | 0.53 |
| optee_os | 0.41 | 0.42 | 0.42 | 0.42 |
| php-src | 0.56 | 0.60 | 0.50 | 0.54 |
| qemu | 0.38 | 0.40 | 0.32 | 0.34 |
| redis | 0.50 | 0.64 | 0.41 | 0.58 |
| RetroArch | 0.52 | 0.64 | 0.40 | 0.56 |
| systemd | 0.42 | 0.41 | 0.32 | 0.31 |
| zephyr | 0.47 | 0.51 | 0.40 | 0.45 |
| zfs | 0.48 | 0.58 | 0.39 | 0.52 |
| **Overall Average** | **0.46** | **0.53** | **0.37** | **0.46** |

Light-GBM Performance comparison for F1 and MCC values between models based on code metrics and models based on HE metrics.

Regarding the Precision-Recall AUC (PR-AUC), models based on Type 2 metrics achieved an average score of 0.87. This result represents a modest but consistent 5.86% increase over the 0.83 average score observed for models based on Type 1 metrics. This trend further reinforces the predictive advantage of the HE-based models across multiple key performance indicators.

TABLE XI
AREA UNDER THE PRECISION-RECALL CURVE FOR LIGHT-GBM

| Project | Type 1 (**AUC**) | Type 2 (AUC) | ↑ (%) |
|---|---|---|---|
| curl | 0.82 | 0.84 | 2.31% |
| emscripten | 0.83 | 0.89 | 6.86 |
| esp-idf | 0.85 | 0.92 | 7.47 |
| FreeRDP | 0.86 | 0.90 | 4.66 |
| git | 0.78 | 0.77 | -1.35 |
| gnucash | 0.77 | 0.84 | 9.01 |
| libuv | 0.74 | 0.90 | 21.16 |
| linux-rbpi | 0.82 | 0.84 | 1.82 |
| linux | 0.77 | 0.78 | 2.21 |
| llvm-project | 0.73 | 0.84 | 14.05 |
| obs-studio | 0.93 | 0.95 | 1.83 |
| openssl | 0.83 | 0.87 | 4.48 |
| openwrt | 0.88 | 0.96 | 8.68 |
| optee_os | 0.90 | 0.91 | 1.78 |
| php-src | 0.89 | 0.92 | 3.82 |
| qemu | 0.81 | 0.82 | 2.09 |
| redis | 0.85 | 0.93 | 9.49 |
| RetroArch | 0.81 | 0.92 | 13.95 |
| systemd | 0.80 | 0.79 | -0.47 |
| zephyr | 0.85 | 0.87 | 2.45 |
| zfs | 0.85 | 0.90 | 6.45 |
| **Overall Average** | **0.83** | **0.87** | **5.85%** |

Light-GBM Performance comparison for AUC values between models based on code metrics and models based on HE metrics.

## B. PERFORMANCE AND IMPORTANCE RANK DISTRIBUTION OF METRICS

Our approach has demonstrated the ability to utilize human factors-based metrics to forecast defect-prone methods based on the PR AUC. The experimental results also show that the prediction power rank distribution of HE-based metrics has the lowest variability of all the models across all projects.

The low variability of prediction importance rank for HE-based features across all projects suggests a reliable and stable underlying relationship between the features and the prediction task. In other words, the observed influence of the feature is not a random occurrence in a single dataset, but rather a consistent and significant contributor to the model's predictions, it provides unique and valuable insights that are not easily replicated by other features, or at least not consistently across all scenarios.

We hypothesize that this observation suggests a more fundamental, potentially causal, relationship between the HE-based features and the prediction task. The features appear to contain substantial information pertinent to the prediction task, which is supported by the finding that combining HE metrics with code and history metrics results in lower average accuracy compared to using HE-based metrics alone.

## C. PRACTICALITY AND ACTIONABILITY

***RQ3: Compared to existing approaches, how do the newly proposed metrics enhance or hinder the explainability, practicality, and actionability of forecasting models as perceived by domain experts?***

In this question we explore how the newly proposed metrics influence the explainability, practicality, and actionability of forecasting models from the perspective of domain experts.

A critical and often overlooked aspect of introducing new analytical tools lies in predicting their adoption and integration into existing workflows. It remains an open and challenging question how domain experts, with their established heuristics and tacit knowledge, will interpret and subsequently act upon the insights generated by these novel metrics. This ambiguity necessitates a rigorous investigation into the mechanisms by which these metrics either enhance or, conversely, hinder the cognitive processes and decision-making frameworks of practitioners. Consequently, this paper directly addresses the currently documented challenges within the literature regarding the operationalization of novel metrics, aiming to reveal the pathways through which these new quantitative measures translate into practical, actionable intelligence for domain experts.

### 1) DIRECTIONS FOR ACTION

To offer more in-depth insight into the importance of our cognitive psychology-based framework, we present a comprehensive summary of the inherent context and insights provided by the novel HE-based metrics in Table XII, framed from a human factors perspective. This foundational understanding directly facilitates the identification of clear directions and actionable strategies for proactively preventing the introduction of defects in software systems.

**Availability of resources:** The presence of up-to-date software requirements, quick reference guides, or knowledge bases can compensate for memory decay by providing readily accessible information.

**Training programs:** The existence of refresher training, drills, simulations, and recurrent practice programs directly addresses memory decay for critical knowledge.

**Work design:** Tasks designed to minimize reliance on perfect recall (e.g., through automation, checklists, highly intuitive interfaces) reduce the impact of memory decay. Commit hooks or similar git automation that warns reviewers of pull requests of potential human error issues in the code. Note that while automation can reduce workload, poorly designed automation can lead to "vigilance decrements" where operators or developers become less alert due to reduced active engagement with the system.

**Culture of learning:** An organization that encourages continuous learning and knowledge sharing can indirectly mitigate decay effects by promoting active engagement with information. Development teams can implement structured review sessions of software requirements and recent source changes to specific components, prior to initiating any code alterations.

**Optimization of work environment:** Supplement natural light with high-quality artificial lighting that mimics natural daylight. To reduce distractions, minimize disruptive noise through acoustic panels, sound-masking systems, or designated quiet zones for focused work. Provide adjustable desks and chairs to promote good posture and reduce physical strain, which can contribute to fatigue.

TABLE XII
CONTEXT AND INSIGHTS OF HE-BASED METRICS

| Metric | Context of the PSF | Directions for action |
|---|---|---|
| **Memory decay** | **Temporal context:** The longer the time since information was last used, the greater the likelihood of memory decay. The longer a method remains untouched, the more probable it is that the developer has forgotten critical details, leading to the introduction of defects upon future modification.<br><br>**Complexity:** Highly complex procedures or knowledge structures might decay differently than simpler, more routine steps.<br><br>**Level of over-learning:** How well the information was initially learned and practiced beyond the point of initial mastery significantly impacts its resistance to decay.<br><br>**Proactive interference:** Older memories interfering with the recall of newer information. For example, if a developer has worked on older versions of a system, recalling the correct procedure for the current version might be hindered by previous knowledge.<br><br>**Retroactive interference:** Newer information interfering with the recall of older information. Learning new aspects of the system might cause the developer to forget details of an older, rarely used component or method.<br><br>**Retrieval:** Memory retrieval is often aided by cues present during encoding. If the context at the time of recall differs significantly from the context of learning, memory decay can seem more pronounced. This may arise, for instance, when a developer revisits a method that has been significantly modified by colleagues since his/her previous involvement. | - Availability of resources<br><br>- Training programs<br><br>- Work design<br><br>- Culture of learning |
| **Alertness** | **Circadian rhythms:** Humans have an internal biological clock that regulates a roughly 24-hour cycle of physiological and behavioral processes, including the sleep-wake cycle, body temperature, and cognitive function.<br><br>**Accumulation of sleep drive:** The longer an individual is awake, the stronger the homeostatic drive to sleep becomes. This "sleep debt" accumulates over time, leading to increasing sleepiness and decreasing alertness.<br><br>**Light exposure:** Bright light, especially blue-spectrum light, is the most potent "zeitgeber" (time-giver) for synchronizing the circadian rhythm. Exposure to bright light in the morning promotes wakefulness, while lack of light in the evening facilitates melatonin production and sleep. Conversely, inadequate lighting in a workspace or excessive light at night can disrupt alertness.<br><br>**Noise and stimuli:** Consistent, monotonous noise or lack of stimulating input can decrease alertness (leading to boredom and vigilance decrements), while sudden or excessive noise can be distracting.<br><br>**Workload and complexity:** Very low or very high workload can both impair alertness. Monotonous programming tasks can lead to boredom and reduced vigilance, while excessively demanding tasks can lead to mental fatigue and burnout.<br><br>**Shift work and irregular schedules:** Working against the natural circadian rhythm (e.g., night shifts, rotating shifts) severely disrupts alertness, leading to chronic sleep deprivation and increased risk.<br><br>**Breaks and rest:** The availability and enforcement of adequate rest breaks are crucial for maintaining alertness during prolonged work periods. | - Optimization of work environment<br><br>- Supporting healthy employee habits<br><br>- Workplace wellness programs<br><br>- Prioritize sleep hygiene<br><br>- Workload management |

Context and insights of metrics derived from human factors theory.

**Supporting healthy employee habits:** Encourage micro-Breaks. Educate employees on the benefits of short, frequent breaks (e.g., 5 minutes every hour) for stretching, standing, or looking away from screens. Provide comfortable, inviting break rooms or outdoor spaces where employees can truly disengage and recharge.

**Workplace wellness programs:** Offer incentives or opportunities for exercise (e.g., discounted gym memberships, on-site fitness classes, walking challenges).

**Prioritize sleep hygiene:** Discourage overtime. Implement policies that encourage a healthy work-life balance and discourage excessive work hours that lead to chronic sleep debt. Provide resources and workshops on the importance of quality sleep and good sleep hygiene practices.

**Workload management:** Managers should set realistic expectations, deadlines and workloads, avoiding chronic overwork that leads to burnout and fatigue. Educate employees on self-monitoring for fatigue and implementing personal countermeasures.

2) ACTIONABILITY

Each of our proposed metrics inherently embodies a rich context, enabling individuals to discern the root causes of the predicted issues and to identify effective corrective measures at individual and organizational level to avert recurring mistakes. We assert the actionability of our metrics, supported by the following reasons:

1. **Specificity:** Our metrics provide specific insights or directions for action, allowing developers and organizations to translate predictions into practical steps such as specific training for improving coding behaviors identified to be lacking.
2. **Relevance:** HE-based metrics align with business goals and decision-making priorities such as Skill Development, Quality Assurance, Compliance and Risk Management, leading to actionable outcomes.
3. **Simplicity:** It is easy to interpret and act upon human factors-based metrics, especially for stakeholders without specialized expertise.
4. **Rich context:** Our metrics consider the broader context and situational factors, producing predictions that are applicable to the current organizational and development process scenario.
5. **Clear impact:** Our metrics, by design, clearly show the potential impact of taking certain actions in the software development process, making it easy to justify and implement solutions. High-risk behaviors suggested by the metric may be avoided, even if they are apparently beneficial, such as extra work hours, doing or avoiding coding tasks at specific times or days, etc.

3) PRACTICALITY

The metrics we introduce are designed with simplicity and accessibility in mind, eliminating the need for intricate interpretation and thereby enhancing their practicality for real-world applications. In this paper, we contend that our metrics exhibit practicality, as evidenced by the following reasons:

1. **Data availability:** Software development is frequently, if not invariably, managed using version control systems like Git, which provide valuable metadata that can be effectively leveraged for analysis. Additionally, the vast array of open-source projects available on platforms such as GitHub offers a rich source of data, enabling extensive opportunities to train and refine our models.
2. **Low cost:** As we leverage software version control system's metadata in our approach, there are no additional financial or resource costs associated with collecting data or implementing the metric.
3. **Technology-independent:** Our metrics can be applied across various languages and technologies without any particular dependencies.
4. **Robustness across different datasets:** Our metrics are based on real-world settings proven by psychology theory. As seen in our experimental results, they consistently rank on top of code and history metrics across different software projects. They are not sensitive to noise or minor variations in data.
5. **Low complexity:** Calculation of our metrics does not require extensive computation, making them practical for real-world applications.

*D. REAL-WORLD APPLICATION OF HE-BASED METRICS*

To substantiate the real-world utility and validity of our proposed metrics, we present compelling evidence of their successful application in identifying defects and vulnerabilities in various projects. This study's development led to the identification of over 30 defects in various software systems through the application of our novel metrics.

Our approach involved defect prediction using human error metrics and code metrics independently. We focused on methods identified as bug-prone solely by the human error metrics. Defects within the selected methods were identified through code reviews and subsequently confirmed via dynamic security testing (fuzzing). Subsequently, we reported our findings to the developers via a responsible disclosure process.

Table XIII presents a selection of some of the key defects discovered in open-source systems via SDP employing HE-based metrics. Git, a tool known for its robust security and high-quality standards, is particularly noteworthy. Despite its strong posture, our model accurately predicted multiple defects within its source code. Additionally, with the help of our model we discovered vulnerabilities in GNU Tar and TagLib that resulted in the assignment of two CVE IDs. Owing to the sensitivity of these systems, some of these issues were not publicly revealed until 2025, just prior to the dissemination of research results.

The predicted defect-prone methods for the findings in Table XIII were consistently overlooked by the model based on code and history metrics, highlighting a significant blind spot in traditional analytical approaches. The underlying data informing these predictions consistently revealed high scores in measures indicative of cognitive failure due to Memory decay and a lack of sustained Alertness among developers.

This correlation between HE scores and actual findings suggests that our metrics uniquely capture the often-subtle human factors contributing to systemic introduction of defects in software systems, thereby offering a powerful new avenue for proactive risk mitigation and system enhancement that transcends the limitations of traditional prediction models based on software metrics.

TABLE XIII
IDENTIFIED SOFTWARE ISSUES IN OPEN-SOURCE SYSTEMS

| Project | Reference | Details |
|---|---|---|
| llvm-project | https://github.com/llvm/llvm-project/issues/74556 | Multiple defects in the llvm infrastructure project. |
| llvm-project | https://github.com/llvm/llvm-project/issues/75048 | Stack overflow in clang. |
| llvm-project | https://github.com/llvm/llvm-project/issues/74726 | LLVM assembler does not properly handles named values in the opaque pointers module. |
| llvm-project | https://github.com/llvm/llvm-project/issues/74732 | LLVM disassembler (llvm-dis) fails to handle unknown attributes. |
| git | https://github.com/git/git/commit/d1bd3a8c3424e818f4117a39fe418909e24cea5f | Out-of-bounds memory reads. |
| git | https://github.com/git/git/commit/dee182941fb685f5d85e61a0e9d97e8e91512f6c | Avoid recursion when unquoting From headers. |
| git | https://github.com/git/git/commit/ba176db511b3438738a4aeb98e574310e697ff5f | Multilpe defects across eleven files when parsing non-bool values. |
| git | https://github.com/git/git/commit/d49cb162fa752d62cf20548ae057471d348e42ae | Handle NULL value when parsing message config. |
| hexdump (util-linux) | https://github.com/util-linux/util-linux/issues/2806 | Heap buffer overflow in Hexdump. |
| dmesg (util-linux ) | https://github.com/util-linux/util-linux/issues/2807 | Multiple memory corruption problems in Dmesg. |
| TagLib | https://github.com/taglib/taglib/issues/1163 | CVE-2023-47466 |
| GNU Tar | https://savannah.gnu.org/bugs/?62387 | CVE-2022-48303 |

Summary of some of the defects detected in open-source systems.

### E. CONCLUSION

In this paper we have outlined our approach to building, training, testing and evaluating HE-based SDP models at the method level. We have introduced a framework based on human factors theory that facilitates the creation of human error-based SDP metrics. We then presented the selection of human error metrics, code metrics and historical measures utilized in SDP. Subsequently, we explained how we compute the metrics and how labeling of methods with defects is done. Afterwards, we employed a supervised machine learning (ML) algorithm to construct the prediction models. We then assessed the performance of these models and the importance of the metrics in predicting defects at the method level. Finally, we presented multiple findings of defects and security issues in open-source systems in the real world, and discussed the actionability and practicality of our approach.

The primary contributions of this study are summarized in the following observations:

1. This study empirically demonstrates the utility of cognitive pyschology theory in the design of software defect prediction (SDP) models, giving a way for practitioners to build derived prediction models based on human factors theory.
2. This research provides empirical evidence that SDP models that leverage human error-based metrics outperform state-of-the-art models, which rely on code and history metrics.
3. Furthermore, this study analyzes the importance of human factors-based metrics in prediction model outputs, revealing their consistently high average importance rankings and robust performance across diverse software projects, even in the presence of noise and minor data variations.
4. The findings underscore the critical role of human factors in defect causation, corroborated by real-world defect and vulnerability discoveries in open-source software systems. The novel metrics contain substantial information pertinent to the prediction task, allowing the models to account for factors influencing human performance and behavior during code development activities. As a result, the novel metrics offer inherent explanations for defect origins and guide preventative strategies. Collectively, this work addresses longstanding challenges in the explainability, practicality, and actionability of software defect prediction.